\documentclass[seceq]{ptptex}
\usepackage{latexsym}
\usepackage{amssymb}
\usepackage{amsfonts}
\usepackage{amsmath}
\usepackage{bm}
\usepackage[dvips]{graphicx}
\usepackage[usenames]{color}
\usepackage{ulem}
\usepackage{multirow}
\allowdisplaybreaks
\renewcommand{\a}{\alpha}
\renewcommand{\b}{\beta}
\renewcommand{\c}{\gamma}

\newcommand{\e}{\epsilon}

\newcommand{\m}{\mu}
\newcommand{\n}{\nu}

\newcommand{\z}{\omega}
\newcommand{\G}{\Gamma}

\title{Gravitational waves from a particle in circular orbits around a Schwarzschild black hole to the 22nd post-Newtonian order}
\author{Ryuichi Fujita$^{1,2}$
}
\inst{
${}^1$Departament de F\'isica, Universitat de les Illes Balears, \\
Cra.\ Valldemossa Km.\ 7.5, Palma de Mallorca, E-07122 Spain
\\
${}^2$Raman Research Institute, Bangalore 560 080, India
}
\abst{
We extend our previous results of the 14th post-Newtonian (PN) order 
expansion of gravitational waves for a test particle in circular orbits 
around a Schwarzschild black hole to the 22PN order, 
i.e. $v^{44}$ beyond the leading Newtonian 
approximation where $v$ is the orbital velocity of a test particle. 
Comparing our 22PN formula for the energy flux with high precision 
numerical results, we find that the relative error of the 22PN flux at 
the innermost stable circular orbit is about $10^{-5}$. 
We also estimate the phase difference between 
the 22PN waveforms and numerical waveforms after a two-year inspiral. 
We find that the dephase is about $10^{-9}$ for $\mu/M=10^{-4}$ and $10^{-2}$ 
for $\mu/M=10^{-5}$ where $\mu$ is the mass of the compact object 
and $M$ the mass of the central supermassive black hole. 
Finally, we construct a hybrid formula of the energy flux 
by supplementing the 4PN formula of the energy flux for circular and 
equatorial orbits around a Kerr black hole with all the present 22PN terms 
for the case of a Schwarzschild black hole. 
Comparing the hybrid formula with the the full numerical results, 
we examine the performance of the hybrid formula for the case of 
Kerr black hole. 
}
\begin{document}
\maketitle
%%%%%%%%%%%%%%%%%%%%%%%%%%%%%%%%%%%%%%%%%%%%%%%%%%%%%%%%%%%%%%%%%%%%%%%%%%%
\section{Introduction} 
\label{sec:intro}
%%%%%%%%%%%%%%%%%%%%%%%%%%%%%%%%%%%%%%%%%%%%%%%%%%%%%%%%%%%%%%%%%%%%%%%%%%%
Gravitational waves (GWs) are expected to be detected in this decade 
by upcoming ground based detectors such as Advanced LIGO~\cite{LIGO}, 
VIRGO~\cite{VIRGO} and KAGRA~\cite{KAGRA}. 
One of the most promising sources of the GWs for the ground based detectors 
is the coalescing stellar-mass compact object binary system. 
For the detection of the GWs and the subsequent parameter estimation of 
the source, one will correlate the noisy signal of the detector with 
the template bank of theoretical waveforms. Thus, it is important to 
predict the waveforms very accurately and efficiently 
not only for developing the data analysis strategy 
but also for extracting the physical information of the source. 

A conventional method to predict inspiral waveforms from coalescing binaries 
is the post-Newtonian (PN) expansion of 
the Einstein equations~\cite{post_Newton}, in which the relative velocity of 
the binary $v$ is assumed to be smaller than the speed of light. 
Currently, the amplitude (orbital phase) of gravitational waves is derived 
up to 3PN (3.5PN), i.e. $v^6$ ($v^7$) 
beyond the leading order~\cite{DJS01,BDE04,BFIJ02,BDEI04,K07,BFIS08,Favata09}, 
for the non-spinning compact binaries in quasi-circular orbits. 
(Note that the 3.5PN amplitude for $\ell=m=2$ mode is derived in 
Ref.~\citen{FMBI2012}.) 
Although the PN expansion accurately describes the early phase of the inspiral, 
the approximation breaks down in the late phase of the inspiral. 
For the late inspiral and the subsequent merger and ringdown phases 
one needs to calculate the waveforms using a full numerical solution of 
the Einstein equations~\cite{Hannam09,Hinder10}. 
Since the computational cost to perform 
full numerical simulations which include the whole process of 
the coalescence 
is too expensive, one should make theoretical templates by 
matching the PN waveforms in the early inspiral phase with 
the numerical waveforms in the late inspiral and the subsequent merger 
and ringdown phases~\cite{EOBTemp2007,PhenomTemp2008}. 
Thus, from the point of view of the computational cost 
it is important to obtain higher PN order expressions of the GWs 
and to investigate whether the high PN order expressions extend 
the region where the PN expansions of GWs are sufficiently reliable. 

For this purpose, we derive the high PN order expressions 
of the GWs focusing on 
a binary system which consists of a compact object of mass $\mu$ 
in circular orbits around a Schwarzschild black hole of mass $M$ by 
assuming $\mu \ll M$. Such extreme mass ratio inspirals (EMRIs) are 
one of the main candidates of gravitational waves for space-based detector 
such as eLISA~\cite{eLISA}. Another approach to model EMRIs is the black hole 
perturbation formalism, which uses the mass ratio as an expansion parameter. 
Although the black hole perturbation theory is limited to the case $\mu \ll M$, 
it is rather easier to compute the high PN order expansions 
of the GWs than using 
the standard PN approximation. Moreover, one can use the numerical results of 
the black hole perturbation theory to investigate the relative accuracy of 
the PN expansions since in the black hole perturbation theory 
there are no assumptions on the orbital velocity of the compact object. 

Using the first order black hole perturbation theory, 
the gravitational waveforms and energy flux to infinity 
for a test particle in circular orbits around a Schwarzschild black hole 
were computed up to 1.5PN in Ref.~\citen{PoissonSch1.5pn}. 
(See Refs.~\citen{PPV2011,LB2009,ref:Thornburg2011} for recent reviews for 
orbital evolution due to the small body's interaction with 
its own gravitational field, i.e. gravitational self-force.) 
The calculation of the energy flux is extended up to 2.5PN 
by numerical fitting in Ref.~\citen{CutlerSch3pn}. 
Then, in Ref.~\citen{TN_log} the 4PN expressions of the energy flux 
were derived by fitting with a very accurate numerical calculation of 
the energy flux. It was also found that $\log v$ terms appear 
at 3PN and 4PN in Ref.~\citen{TN_log}. 
These $\log v$ terms were confirmed in Ref.~\citen{TS}, 
which computed analytically the 4PN expressions of the energy flux. 
These calculations were extended to 5.5PN for the energy flux in 
Ref.~\citen{ref:TTS} and for the waveforms in Ref.~\citen{FI2010}. 
Recently the 14PN gravitational waveforms for a test particle in circular 
orbits around a Schwarzschild black hole have been derived~\cite{14PN}. 
By comparing the 14PN energy flux with numerical results, 
it is shown that the relative error of the post-Newtonian energy flux 
becomes smaller when the PN order is higher. 
It is also shown that the 14PN waveforms using 
factorized resummation suggested in Ref.~\citen{DIN} will provide 
the data analysis performances of EMRIs comparable to the ones resulting from 
high precision numerical waveforms. 
In this paper we extend the 14PN results in Ref.~\citen{14PN} to the 22PN, 
which is the highest order we can derive with reasonable time 
using our current code.\footnote{
We note that the number of terms necessary to derive the PN expressions 
grows exponentially when the PN order becomes higher. 
Our current code uses $70$, $3.3\times 10^{2}$, $1.9\times 10^{3}$, 
$1.1\times 10^{4}$ and $3.3\times 10^{4}$MBytes memory, 
taking seconds to a few days, to compute multipolar waveforms for 
$\ell=m=2$ mode at 6PN, 10PN, 14PN, 18PN and 22PN respectively. 
Thus, it is difficult to obtain 23PN or higher order expressions 
with reasonable time by using our current code.} 
We find that if one does not use any resummation technique 
the 22PN gravitational waveforms will be necessary to 
achieve the data analysis accuracies comparable to the ones 
using high precision numerical waveforms. 

This paper is organized as follows. 
In Sec.~\ref{sec:formula}, we start with a brief summary of 
the first order black hole perturbation formalism. 
Sec.~\ref{sec:MST} describes a brief summary of an analytic formalism to 
compute homogeneous solutions of the Teukolsky equation. 
Sec.~\ref{sec:results} is devoted to the comparison of 
our 22PN expressions with high precision numerical results: 
The total energy flux to infinity is compared in Sec.~\ref{sec:flux}, 
the phase of gravitational waveforms during two-year inspiral 
Sec.~\ref{sec:dephase} and the total energy flux to infinity 
for the case of a Kerr black hole Sec.~\ref{sec:flux_kerr}. 
We conclude with a brief summary in Sec.~\ref{sec:summary}. 
Since the 22PN expressions are too large to be shown in the paper, 
we only show the 7PN expression 
of the total energy flux to infinity in Appendix~\ref{sec:7pn_formula}. 
In Appendix~\ref{sec:22pn_formula} we show the 22PN expression of 
the total energy flux to infinity, which can be used for numerical 
computation in double precision. The 22PN expressions for 
the modal energy flux to infinity and factorized waveforms~\cite{DIN} 
will be publicly available online~\cite{BHPC}. 
Throughout this paper, we work in the units of $c=G=1$. 
%%%%%%%%%%%%%%%%%%%%%%%%%%%%%%%%%%%%%%%%%%%%%%%%%%%%%%%%%%%%%%%%
\section{Teukolsky formalism}
\label{sec:formula}
%%%%%%%%%%%%%%%%%%%%%%%%%%%%%%%%%%%%%%%%%%%%%%%%%%%%%%%%%%%%%%%%%%%%%%%%%%%
We consider the gravitational waves emitted by a test particle 
moving on circular orbits around a Schwarzschild black hole 
using the Teukolsky formalism~\cite{Teukolsky:1973ha}. 
In this section, we recapitulate necessary equations 
following the notation in Ref.~\citen{ST}. 

In the Teukolsky formalism, the gravitational perturbation of a Schwarzschild 
black hole is represented by the Weyl scalar $\Psi_4$, which 
represents the gravitational waves going out to infinity as 
\begin{eqnarray}
\Psi_4\rightarrow\frac{1}{2}(\ddot{h}_{+}-i\,\ddot{h}_{\times }),\,\,\,{\rm for}\,\,\,r\rightarrow\infty, 
\end{eqnarray}
where dot $\dot{\,}$ represents a time derivative, $d/dt$.

The Weyl scalar can be separated into radial and angular parts if we
decompose $\Psi_4$ in Fourier harmonic components as
\begin{eqnarray}
\Psi_4=\frac{1}{r^4}\displaystyle \sum_{\ell,m}\int_{-\infty}^{\infty} d\omega 
e^{-i\omega t} R_{\ell m\omega}(r)
\ _{-2}Y_{\ell m}(\theta,\varphi),
\end{eqnarray}
where $\ell\ge 2$, $-\ell\le m\le\ell$ and $_{s}Y_{\ell m}(\theta,\varphi)$ 
is the spin-weighted spherical harmonics~\cite{K07}. The radial function 
$R_{\ell m\omega}(r)$ satisfies the inhomogeneous Teukolsky equation, 
\begin{equation}
 \left[\Delta^2{d\over dr}\left({1\over \Delta}{d\over dr}\right)+U(r)\right]
  R_{\ell m\omega}(r) = T_{\ell m\omega}(r),
\label{eq:Teu}
\end{equation}
with $\Delta=r(r-2M)$ and 
\begin{equation}
 U(r)={r^2\over\Delta}\left[\omega^2 r^2
-4i\omega(r-3M)\right]-(\ell-1)(\ell +2), 
\end{equation}
where $T_{\ell m\omega}$ is the source term which is constructed from 
the energy momentum tensor of the small particle. 

We use the Green function method to solve the inhomogeneous 
Teukolsky equation Eq.~(\ref{eq:Teu}). 
The outgoing-wave solution of the Teukolsky equation Eq.~(\ref{eq:Teu}) 
at infinity is given by 
\begin{align}
 R_{\ell m\omega}(r\rightarrow\infty)  &= {r^3 e^{i\omega r^{*}}\over 
    2i\omega B_{\ell m\omega}^{\rm inc}}\int_{2M}^{\infty} dr {R_{\ell m\omega}^{\rm in} T_{\ell m\omega}(r)\over \Delta^{2}},\cr
   &\equiv r^3 e^{i\omega r^{*}}\tilde Z_{\ell m\omega}, 
\label{eq:Zinf_b}
\end{align}
where $r^{*}=r+2M\ln(r/2M -1)$ and 
$R_{\ell m\omega}^{in}$ is the homogeneous solution of Eq.~(\ref{eq:Teu}) 
which satisfies the ingoing-wave boundary condition at horizon as 
\begin{equation}
R_{\ell m\omega}^{\rm in}=\left\{
  \begin{array}{lcc}
    B_{\ell m\omega}^{\rm trans}\Delta^2 e^{-i\omega r^{*}} & \hbox{for} 
    & r^{*}\rightarrow -\infty, \\
    r^3 B_{\ell m\omega}^{\rm ref} e^{i\omega r^{*}} +
    r^{-1} B_{\ell m\omega}^{\rm inc}  e^{-i\omega r^{*}} & \hbox{for} 
    & r^{*}\rightarrow +\infty.
   \end{array}
\right.
\label{eq:Rin_asymp}
\end{equation}

When the test particle moves on a circular orbit around a Schwarzschild 
black hole, the angular frequency $\Omega$, the specific energy $\tilde E$ 
and the angular momentum $\tilde L$ of the particle are given by 
\begin{equation}
\Omega=\sqrt{\frac{M}{r_0^3}},\,\,\,
\tilde E=\frac{r_{0}-2M}{\sqrt{r_0(r_0-3M)}},\,\,\,
\tilde L=\frac{\sqrt{M r_{0}}}{\sqrt{1-3M/r_0}},
\end{equation}
where $r_0$ is the orbital radius. 
The frequency spectrum of $T_{\ell m\omega}$ has peaks at the harmonics of 
the orbital frequency $\omega=m\Omega$. Then one finds that 
$\tilde Z_{\ell m\omega}$ in Eq.~(\ref{eq:Zinf_b}) takes the form 
\begin{equation}
 \tilde Z_{\ell m\omega}=Z_{\ell m\omega}\delta(\omega-m\Omega),
\label{eq:tildeZ}
\end{equation}
where
\begin{eqnarray}
Z_{\ell m\omega}& = &{\m\pi\over i\omega (r_0/M)^2 B_{\ell m\omega}^{{\rm inc}}}
  \Biggl[\Biggl\{
      -_0 b_{\ell m} -2\,i\, _{-1}b_{\ell m}
             \left(1+{i\,\omega\,r_0^2\over 2\,(r_0-2M)}\right)\cr &&
    + i\, _{-2} b_{\ell m}{\omega\,r_0\over (1-2M/r_0)^{2}}
     \left(1-\frac{M}{r_0}+{1\over 2}i\,\omega\,r_0\right)\Biggr\} R_{\ell m\omega}^{{\rm in}}
\cr &&
    +\left\{i\,_{-1}b_{\ell m}-_{-2}b_{\ell m}
         \left(1+{i\,\omega\,r_0^2\over r_0-2M}\right)\right\}
           r_0\, {R^{{\rm in}}_{\ell m\omega}}'(r_0)\cr &&
      +{1\over 2} {}_{-2} b_{\ell m} \,r_0^2\, {R_{\ell m\omega}^{{\rm in}}}''(r_0)
      \Biggr],
\label{eq:Z8q0e0}
\end{eqnarray}
and prime, $'$, denotes differentiation with respect to $r$ and 
$_s b_{\ell m}$ are defined by 
\begin{subequations}
\begin{align}
_0 b_{\ell m}  = & {1\over2}\left[(\ell-1)\ell(\ell+1)(\ell+2)\right]^{1/2}%\cr&
{}_{0} Y_{\ell m}\left({\pi\over 2},\,0\right){\tilde E r_0\over r_0-2M},
\\
_{-1} b_{\ell m}  = & \left[(\ell-1)(\ell+2)\right]^{1/2}
    {}_{-1} Y_{\ell m}\left({\pi\over 2},\,0\right)\frac{\tilde L}{r_0},
\\
_{-2} b_{\ell m}  = & _{-2} Y_{\ell m}\left({\pi\over 2},\,0\right)
            \tilde L\Omega. 
\end{align}
\label{eq:blm}
\end{subequations}

Using the amplitudes $Z_{\ell m\omega}$, the gravitational wave 
luminosity is given by 
\begin{equation}
{dE\over dt}=\sum_{\ell=2}^{\infty} \sum_{m=-\ell}^{\ell}
   \frac{\vert Z_{\ell m\omega}\vert^2}{4\pi \omega^2}. 
\label{eq:flux}
\end{equation}
Choosing $(\theta,\varphi)$ as the angles defining the location of 
the observer relative to the source, the gravitational waveforms can be 
expressed as 
\begin{align}
h_{+}-i\,h_{\times }
=-\frac{2}{r}\,\sum _{\ell,m} \frac{Z_{\ell m\omega}}{\omega^2}\,_{-2}Y_{\ell m}(\theta,\varphi )\,e^{i \omega (r^{*}-t)},
\label{eq:wave}
\end{align}
where $\omega=m\Omega$ is the frequency of the gravitational waves. 

We derive the post-Newtonian expansions of the gravitational wave luminosity 
Eq.~(\ref{eq:flux}) and gravitational waveforms Eq.~(\ref{eq:wave}) by 
computing the amplitude $Z_{\ell m\omega}$. 
For the computation of $Z_{\ell m\omega}$, one needs to obtain 
the series expansion of the ingoing-wave Teukolsky function 
$R_{\ell m\omega}^{{\rm in}}$ in terms of 
$\epsilon\equiv 2M\omega=2 M m \Omega=O(v^3)$ 
and $z\equiv\omega r=O(v)$ and the asymptotic amplitudes 
$B^{{\rm inc}}_{\ell m\omega}$ in terms of $\epsilon$, where $v=(M/r_0)^{1/2}$. 
We use a formalism developed by Mano, Suzuki and 
Takasugi~\cite{ref:MST,ref:MSTR} to compute $R_{\ell m\omega}^{{\rm in}}$ and 
$B^{{\rm inc}}_{\ell m\omega}$. In the next section, we give a brief review of 
the formalism for the convenience of the reader. 
%%%%%%%%%%%%%%%%%%%%%%%%%%%%%%%%%%%%%%%%%%%%%%%%%%%%%%%%%%%%%%%%%%%%%%%%%%%
\section{Analytic solutions of the homogeneous Teukolsky equation}
\label{sec:MST}
%%%%%%%%%%%%%%%%%%%%%%%%%%%%%%%%%%%%%%%%%%%%%%%%%%%%%%%%%%%%%%%%%%%%%%%%%%%
We adopt the formalism developed by Mano, Suzuki and Takasugi (MST) 
~\cite{ref:MST,ref:MSTR} to obtain high post-Newtonian order expansion 
of gravitational waves. In the MST formalism, 
the homogeneous solutions of the Teukolsky equation 
are expressed using hypergeometric functions around the horizon 
and Coulomb wave functions around infinity. 
Since the matching of the two kinds of solutions can be done analytically 
in the overlapping region of convergence, 
one can obtain analytic expressions of the homogeneous solutions 
without numerical integration. 
Moreover, the series expansions are naturally related to the low frequency 
expansion (See Sec.~\ref{sec:MST_PN}). 
Thus, the formalism is very powerful to compute high post-Newtonian order 
expansion of gravitational waves. 
Using the MST formalism, the energy absorption of gravitational waves 
into the horizon induced by a particle in circular orbits around 
the equatorial plane of a Kerr black hole was calculated 
up to 6.5PN order beyond the quadrupole formula~\cite{TMT}. 
The energy flux of gravitational waves to infinity by a particle in 
slightly eccentric and inclined orbits around a Kerr black hole 
was also computed up to 2.5PN order in Refs.~\citen{STHGN,Ganz}. 
In Refs.~\citen{FI2010} and \citen{spin_resum}, 
the author of this paper was part of collaborations that 
applied the formalism to obtain gravitational waveforms 
for a test particle in circular orbits around a Schwarzschild 
black hole up to 5.5PN and a Kerr black hole up to 4PN. 
Extending the formalism to very high post-Newtonian order, we derived 
the 14PN expressions of gravitational waves for a test particle in circular 
orbits around a Schwarzschild black hole in Ref.~\citen{14PN}. 
Although the MST formalism can be applied to the case of a Kerr black hole, 
we set $q=0$, where $q$ is the non-dimensional spin parameter of the Kerr 
black hole, since we are dealing with the the case of a Schwarzschild 
black hole in this paper. 
We refer the interested reader to a review Ref.~\citen{ST} 
for details of the formalism. 
%%%%%%%%%%%%%%%%%%%%%%%%%%%%%%%%%%%%%%%%%%%%%%%%%%%%%%%%%%%%%%%%%%%%%%%%%%%
\subsection{Ingoing-wave solution}
\label{sec:Coulomb}
%%%%%%%%%%%%%%%%%%%%%%%%%%%%%%%%%%%%%%%%%%%%%%%%%%%%%%%%%%%%%%%%%%%%%%%%%%%
A homogeneous solution of the Teukolsky equation 
in a series of Coulomb wave functions $R_{{\rm C}}^{\n}$ is defined as 
\begin{equation}
R_{{\rm C}}^{\n}=
z\left(1-{\e  \over{{z}}}\right)^{2-i\e} \, 
\displaystyle\sum_{n=-\infty}^{\infty}
(-i)^n\frac{(\n-1-i\e)_n}{(\n+3+i\e)_n} a_n^{\n} F_{n+\n}(2\,i-\e,z),
\label{eq:series of Rc}
\end{equation}
where $z=\z r$, $(a)_{n}=\Gamma(a+n)/\Gamma(a)$, 
and $F_{N}(\eta,z)$ is a Coulomb wave function defined by 
\begin{eqnarray}
F_{N}(\eta,z)&=&e^{-iz}2^{N}z^{N+1}\frac{\G(N+1-i\eta)}{\G(2N+2)}
\Phi(N+1-i\eta,2N+2;2iz), 
\label{eq:defcoulomb}
\end{eqnarray}
where $\Phi(\a,\b;z)$ is a confluent hypergeometric function~\cite{handbook}. 
Observe that the so-called renormalized angular momentum $\nu$ is introduced 
in the homogeneous solution in a series of Coulomb wave functions 
Eq.~(\ref{eq:series of Rc}). 
This parameter $\nu$ is a generalization of $\ell$ and $\nu\rightarrow\ell$ as 
$\e\rightarrow 0$. 
The parameter is determined by the conditions that 
the series converges and actually represents a homogeneous solution of 
the Teukolsky equation. 

Substituting the solution in a series of Coulomb wave functions 
Eq.~(\ref{eq:series of Rc}) into the Teukolsky equation 
Eq.~(\ref{eq:Teu}) with $T_{\ell m\omega}=0$, 
we obtain a three-term recurrence relation for the 
expansion coefficients $a_{n}^{\n}$
\begin{eqnarray}
\alpha_n^\nu a_{n+1}^{\n}+\beta_n^{\nu} a_{n}^{\n}+\gamma_n^\nu a_{n-1}^{\n}=0,
\label{eq:3term}
\end{eqnarray}
where
\begin{subequations}
\begin{eqnarray}
\a_n^\n&=&{i\e (n+\n-1+i\e)(n+\n-1-i\e)(n+\n+1+i\e)
\over{(n+\n+1)(2n+2\n+3)}},
\\
\b_n^\n&=&-\ell(\ell+1)+(n+\n)(n+\n+1)+2\e^2
+{\e^2 (4+\e^2) \over{(n+\n)(n+\n+1)}},\\
\c_n^\n&=&-{i\e  (n+\n+2+i\e)(n+\n+2-i\e)(n+\n-i\e)
\over{(n+\n)(2n+2\n-1)}}.
\end{eqnarray}
\label{eq:3term_abc}
\end{subequations}

The series Eq.~(\ref{eq:series of Rc}) converges if 
the renormalized angular momentum $\nu$ satisfies 
the following equation~\cite{ref:MST,ST}, 
\begin{eqnarray}
R_{n}L_{n-1}=1,
\label{eq:consistency}
\end{eqnarray}
where $R_{n}$ and $L_{n}$ are defined by continued fractions, 
which are derived by the three-term recurrence relation Eq.(\ref{eq:3term}), 
as 
\begin{eqnarray}
R_n&\equiv& {a_{n}^{\n}\over a_{n-1}^{\n}}
=-{\gamma_n^\nu\over {\beta_n^\nu+\alpha_n^\nu R_{n+1}}}%,\cr&=&
=-{\gamma_{n}^\nu\over \beta_{n}^\nu-}\,
{\alpha_{n}^\n\gamma_{n+1}^\nu\over \beta_{n+1}^\nu-}\,
{\alpha_{n+1}^\n\gamma_{n+2}^\nu\over \beta_{n+2}^\nu-}\cdots,
\label{eq:Rncont}\\
L_n&\equiv& {a_{n}^{\n}\over a_{n+1}^{\n}}
=-{\alpha_n^\nu\over {\beta_n^\nu+\gamma_n^\nu L_{n-1}}}%,\cr&=&
=-{\alpha_{n}^\nu\over \beta_{n}^\nu-}\,
{\alpha_{n-1}^\n\gamma_{n}^\nu\over \beta_{n-1}^\nu-}\,
{\alpha_{n-2}^\n\gamma_{n-1}^\nu\over \beta_{n-2}^\nu-}\cdots.
\label{eq:Lncont}
\end{eqnarray}
From the continued fractions, $R_{n}$ and $L_{n}$, one can obtain 
two kinds of the expansion coefficients, $a_{n}^{\n}$. 
In general, these two kinds of the expansion coefficients do not coincide. 
Eq.~(\ref{eq:consistency}) is a condition such that the two types of 
the expansion coefficients coincide and the series converges~\cite{ref:MST,ST}. 
If we choose $\nu$ using Eq.~(\ref{eq:consistency}), 
the series of Coulomb wave function Eq.~(\ref{eq:series of Rc}) 
converges for $r>r_{+}$. 
From Eq.~(\ref{eq:3term_abc}), we find $\a_{-n}^{-\n-1}=\c_{n}^{\n}$ 
and $\b_{-n}^{-\n-1}=\b_{n}^{\n}$ so that 
$a_{-n}^{-\n-1}$ satisfies the same recurrence relation Eq.~(\ref{eq:3term}) 
as $a_{n}^{\n}$ does. 
This shows that $R_{{\rm C}}^{-\n-1}$ is also a homogeneous solution 
of the Teukolsky equation, which converges for $r>r_{+}$.

As for the homogeneous solution in a series of hypergeometric functions, 
which converges for $r<\infty$, 
the expansion coefficients satisfy the same three-term 
recurrence equation~(\ref{eq:3term}) derived by using a series of 
Coulomb wave functions. 
Thus one can use the same renormalized angular momentum $\nu$ to compute 
both series. This fact is important to match the solution in a series of 
Coulomb wave functions, which converges at infinity, with the one 
in a series of hypergeometric functions, which converges at the horizon. 
As a result of the matching, 
one can obtain the ingoing-wave solution $R_{lm\omega}^{{\rm in}}$, 
which converges in the entire region, as 
\begin{eqnarray}
R_{lm\omega}^{{\rm in}}=K_{\n}R_{{\rm C}}^{\n}+K_{-\n-1}R_{{\rm C}}^{-\n-1}, 
\label{eq:secondRin}
\end{eqnarray}
where 
\begin{eqnarray}
K_{\n}
&=&\frac{e^{i\e}(2\e)^{-2-\n-N}2^{2}i^{N}\Gamma(3-2i\e)\Gamma(N+2\n+2)}
{\Gamma(N+\n+3+i\e)\Gamma(N+\n+1+i\e)\Gamma(N+\n-1+i\e)}\nonumber \\
&&\times\left(\sum_{n=N}^{\infty}(-1)^{n}\frac{\Gamma(n+N+2\n+1)}{(n-N)!}
\frac{\Gamma(n+\n-1+i\e)\Gamma(n+\n+1+i\e)}
{\Gamma(n+\n+3-i\e)\Gamma(n+\n+1-i\e)}a_{n}^{\n}\right)\nonumber \\
&&\times\left(\sum_{n=-\infty}^{N}\frac{(-1)^{n}}{(N-n)!(N+2\n+2)_{n}}
\frac{(\n-1-i\e)_{n}}{(\n+3+i\e)_{n}}a_{n}^{\n}\right)^{-1},
\end{eqnarray}
and $N$ can be any integer. 
The factor $K_\n$ is a constant to match the solutions 
in the overlap region of convergence and 
independent of the choice of $N$.

One can obtain analytic expressions for the asymptotic amplitudes 
$B^{{\rm trans}}_{lm\omega}$, $B^{{\rm inc}}_{lm\omega}$ and $B^{{\rm ref}}_{lm\omega}$ 
defined in Eq.~(\ref{eq:Rin_asymp}) by comparing the asymptotic behavior of 
$R_{lm\omega}^{{\rm in}}$ Eq.~(\ref{eq:Rin_asymp}) 
with Eq.~(\ref{eq:secondRin}) at $r^{*}\rightarrow\pm\infty$. 
They are derived as 
\begin{subequations}
\begin{eqnarray}
B^{{\rm trans}}_{lm\omega}
&=&\left(\frac{\e}{\omega}\right)^{-4}
\sum_{n=-\infty}^{\infty}a_{n}^{\n},\\
B^{{\rm inc}}_{lm\omega}
&=&\omega^{-1}\left[K_{\n}-ie^{-i\pi\n}
\frac{\sin\pi(\n+i\e)}{\sin\pi(\n-i\e)}K_{-\n-1}\right]%\cr&&\times 
A_{+}^{\n}e^{-i\e\ln\e},\\
B^{{\rm ref}}_{lm\omega}
&=&\omega^{3}[K_{\n}+ie^{i\pi\n}K_{-\n-1}]A_{-}^{\n}e^{i\e\ln\e},
\end{eqnarray}
\label{eq:asymp_amp}
\end{subequations}
where
\begin{subequations}
\begin{eqnarray}
A_{+}^{\n}&=&2^{-3-i\e}e^{-\frac{\pi\e}{2}}e^{\frac{\pi}{2}i(\n+3)}
\frac{\G(\n+3+i\e)}{\G(\n-1-i\e)}%\cr&&\times
\sum_{n=-\infty}^{+\infty}a_{n}^{\n},
\label{eq:Ap}
\\
A_{-}^{\n}&=&2^{1+i\e}e^{-\frac{\pi\e}{2}}e^{-\frac{\pi}{2}i(\n-1)}%\cr&&\times
\sum_{n=-\infty}^{+\infty}(-1)^{n}\frac{(\n-1-i\e)_{n}}{(\n+3+i\e)_{n}}a_{n}^{\n}.
\end{eqnarray}
\end{subequations}

We summarize how to compute the gravitational waves using 
the MST formalism. One first should determine $\n$ by solving 
Eq.~(\ref{eq:consistency}). Then, 
one can compute the expansion coefficients $a_n^{\n}$ using 
the continued fractions Eq.~(\ref{eq:Rncont}) for $n>0$ 
and Eq.~(\ref{eq:Lncont}) for $n<0$ 
with the condition $a_0^{\n}=a_0^{-\n-1}=1$. 
Substituting the expansion coefficients $a_n^{\n}$ into 
Eqs~(\ref{eq:secondRin}) and (\ref{eq:asymp_amp}), 
one can compute the ingoing-wave solution of the Teukolsky 
equation $R_{\ell m\omega}^{{\rm in}}$ 
and the asymptotic amplitudes $B_{\ell m\omega}^{{\rm inc}}$, 
which are necessary ingredients to compute $Z_{\ell m\omega}$ 
in Eq.~(\ref{eq:Z8q0e0}) and hence allow one to compute 
the energy flux to infinity Eq.~(\ref{eq:flux}) and 
gravitational waveforms Eq.~(\ref{eq:wave}). 
%%%%%%%%%%%%%%%%%%%%%%%%%%%%%%%%%%%%%%%%%%%%%%%%%%%%%%%%%%%%%%%%%%%%%%%%%%%
\subsection{Low frequency expansions of solutions}
\label{sec:MST_PN}
%%%%%%%%%%%%%%%%%%%%%%%%%%%%%%%%%%%%%%%%%%%%%%%%%%%%%%%%%%%%%%%%%%%%%%%%%%%
In the practical calculation to determine $\n$, 
we solve an alternative equation which is equivalent 
to Eq.~(\ref{eq:consistency}) for $n=1$ 
\begin{eqnarray}
\beta_0^{\nu}+\alpha_0^\nu R_{1}+\gamma_0^\nu L_{-1}=0,
\label{eq:determine_nu}
\end{eqnarray}
where $R_{1}$ and $L_{-1}$ are given by the continued fractions 
Eq.~(\ref{eq:Rncont}) and Eq.~(\ref{eq:Lncont}) respectively. 

In the followings, we consider the low frequency approximation for 
Eq.~(\ref{eq:determine_nu}). 
In the limit of low frequency, we solve Eq.~(\ref{eq:determine_nu}) 
by requiring $\nu\rightarrow\ell$ for $\e\rightarrow 0$. 
Since the orders of $\a_n^\n$, $\c_n^\n$, $\b_n^\n$, 
$R_{1}$ and $L_{-1}$ in $\e$ are $O(\e)$, $O(\e)$, $O(1)$, 
$O(\e)$ and $O(\e)$ respectively 
except for certain values of $n<0$~\cite{ref:MST,ST}, 
one can systematically compute the low frequency expansion of $\nu$. 
The closed analytic form of $\n$ at $O(\e^2)$ is given in 
Refs.~\citen{ref:MST,ST}. 
The expansion coefficients $a_n^{\n}$ can be computed by using 
$R_{n}$ and $L_{-|n|}$, whose order in $\e$ are $O(\e)$ for sufficiently 
large $n$. Therefore, the order of the expansion coefficients $a_n^{\n}$ 
in $\e$ are $O(\e^{|n|})$ for sufficiently large $|n|$. 
Using the low frequency expansion of $a_n^{\n}$, one can easily derive 
that of the asymptotic amplitudes Eq.~(\ref{eq:asymp_amp}). 
Since the homogeneous solution of the Teukolsky equation 
$R_{\ell m\omega}^{{\rm in}}$ Eq~(\ref{eq:secondRin}) is a function of $z=O(v)$ 
and $a_n^{\n}\sim O(\e^{|n|})=O(v^{3\,|n|})$, the homogeneous solution 
is naturally related to 
the post-Newtonian approximation. For the calculation of 
$R_{\ell m\omega}^{{\rm in}}$ in the post-Newtonian approximation, however, 
one should add a larger number of terms in $R_{{\rm C}}^{\n}$ 
Eq.~(\ref{eq:series of Rc}) for $n<0$ than that for $n>0$. 
This is because the post-Newtonian order of a series of 
Coulomb wave functions for $n<0$, 
$(a_{-|n|}^{\n}F_{-|n|+\n}(2\,i-\e,z))/(a_0^{\n}F_{\n}(2\,i-\e,z))\sim O(v^{2\,|n|})$, 
grows slower than that for $n>0$, 
$(a_{n}^{\n} F_{n+\n}(2\,i-\e,z))/(a_0^{\n} F_{\n}(2\,i-\e,z))\sim O(v^{4\,n})$
~\cite{FI2010}. Then, it is straightforward to compute the post-Newtonian 
expansion of the energy flux to infinity Eq.~(\ref{eq:flux}) and 
gravitational waveforms Eq.~(\ref{eq:wave}) using the post-Newtonian 
expansion of $Z_{\ell m\omega}$ Eq.~(\ref{eq:Z8q0e0}). 
The closed analytic form of $Z_{\ell m\omega}$ at 2.5PN 
for the case of a test particle moving in circular orbits around a 
Schwarzschild black hole is given in Ref.~\citen{FI2010}. 
%%%%%%%%%%%%%%%%%%%%%%%%%%%%%%%%%%%%%%%%%%%%%%%%%%%%%%%%%%%%%%%%
\section{Comparison with numerical results}
\label{sec:results}
%%%%%%%%%%%%%%%%%%%%%%%%%%%%%%%%%%%%%%%%%%%%%%%%%%%%%%%%%%%%%%%%
\subsection{Total energy flux to infinity}
\label{sec:flux}
\begin{figure}[htbp]
\begin{center}
\includegraphics[width=70mm]{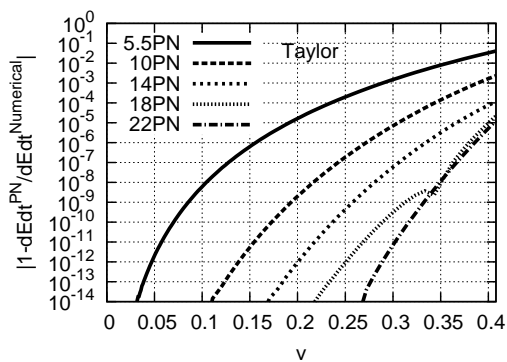}
\end{center}
\caption{\label{fig:flux}Absolute values of the difference of the energy flux 
to infinity between numerical results and 
the PN approximation as a function of the orbital velocity. Note that 
the relative error of the energy flux between 22PN and the numerical 
calculation is an order of magnitude smaller than the one between 
14PN and the numerical calculation even around ISCO, $v=1/\sqrt{6}=0.40825$. 
}
\end{figure}

%%%%%%%%%%%%%%%%%%%%%%%%%%%%%%%%%%%%%%%%%%%%%%%%%%%%%%%%%%%%%%%%
Fig.~\ref{fig:flux} shows the comparison of the energy flux to infinity between 
a high precision numerical calculation and post-Newtonian approximations 
up to 22PN. The numerical computation is based on a technique in 
Ref.~\citen{FT1,FT2}. The accuracy of the numerical calculation is mainly 
limited by truncation of $\ell$-mode. In this work we set $\ell=25$ which 
gives relative error better than $10^{-14}$ up to the innermost stable 
circular orbit (ISCO). $n$-PN flux needs $\ell$ up to $n+2$. 
We note that the agreement between the numerical energy flux and 
post-Newtonian energy flux becomes better when the PN order is higher 
even around ISCO. The relative error of the 22PN energy flux around ISCO 
is about $10^{-5}$ and an order of magnitude better than that of 
14PN energy flux, which was derived in our previous paper~\cite{14PN}. 
We also note that the accuracy can be improved if we compute the energy 
flux using factorized resummation waveforms~\cite{DIN}. 
If one uses the factorized resummation waveforms, 
the relative error of the energy flux around ISCO can be reduced to 
$10^{-6}$ for 14PN expressions~\cite{14PN} and $10^{-7}$ 
for our 22PN expressions. 

%%%%%%%%%%%%%%%%%%%
%%%%%%%%%%%%%%%%%%%%%%%%%%%%%%%%%%%%%%%%%%%%%%%%%%%%%%%%%%%%%%%%
\subsection{Phase difference during two-year inspiral}
\label{sec:dephase}
%%%%%%%%%%%%%%%%%%%%%%%%%%%%%%%%%%%%%%%%%%%%%%%%%%%%%%%%%%%%%%%%
From Fig.~\ref{fig:flux}, one may expect that 22PN expression for 
gravitational waves can be used for data analysis of EMRIs 
since the relative error with a high precision numerical calculation 
is about $10^{-5}$ around ISCO. 
To investigate the applicability of the PN expressions to the data analysis, 
we compute the phase difference during two years 
quasi-circular inspiral between the PN waveforms and the numerical waveforms. 
Following Ref.~\citen{EOB_EMRI}, we choose two systems, 
denoted as System-I (System-II), which correspond to 
the early (late) inspiral phase of an EMRI in the frequency band of eLISA. 
System-I has masses $(M,\m)=(10^5,10)M_{\odot}$ 
and inspirals from $r_0\simeq 29M$ to $r_0\simeq 16M$ with 
associated frequencies $f_{\rm GW}\in [4\times 10^{-3},10^{-2}]$Hz. 
System-II has masses $(M,\m)=(10^6,10)M_{\odot}$ 
and starts inspiral from $r_0\simeq 11M$ to $r_0\simeq 6.0M$ with 
frequencies $f_{\rm GW}\in [1.8\times 10^{-3},4.4\times 10^{-3}]$Hz. 
System-I (II) has the mass ratio of $10^{-4}$ ($10^{-5}$) and 
$\sim 1\times 10^{6}$ ($\sim 5\times 10^{5}$) rads of 
the orbital phase due to the two-year inspiral. 

For the calculation of the phase, we use the method described in 
Ref.~\citen{Hughes2001}, which is also used in Refs.~\citen{EOB_EMRI,FI2010}. 
The phase of the waveforms can be described as 
$m\int_0^{t}\Omega(t')dt'-\Psi_{\ell m}(t)$, 
where $\Omega(t)=\sqrt{M/r_0^3(t)}$ and 
$\Psi_{\ell m}$ is the phase of $Z_{\ell m\omega}$. 
To compute the radius of the orbits as a function of time $r_0(t)$, 
we use a interpolation method to save computational time. 
For the interpolation, using total energy flux $dE/dt$ induced by a particle 
we compute $d r_0/dt$ for $10^3$ points data of $v$ from $v=0.01$ to $v=0.408$. 
Then, from the $10^3$ points data of ($v$, $d r_0/dt$), 
we compute ($r_0(t)$, $\Psi_{\ell m}(t)$) 
using the cubic spline interpolation~\cite{Recipes}. 

\begin{figure}[htbp]
\begin{center}
\includegraphics[width=70mm]{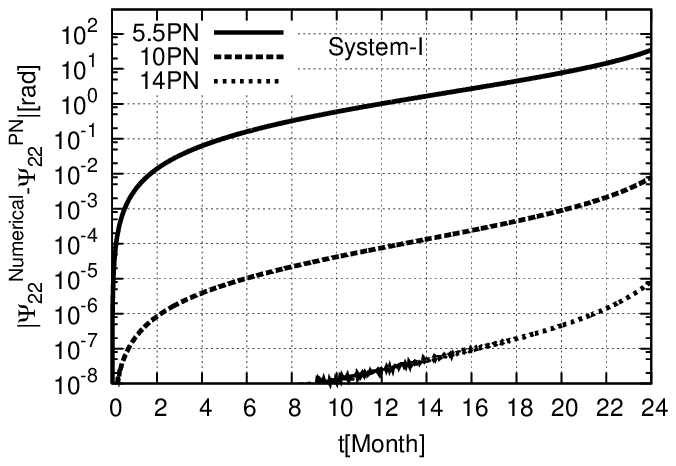}%
\includegraphics[width=70mm]{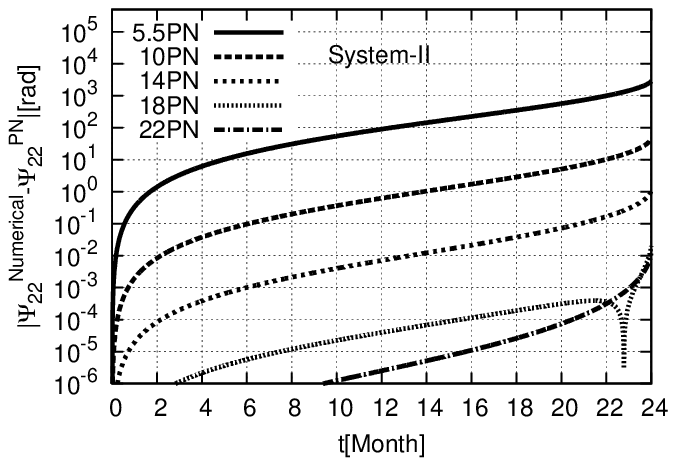}
\caption{Absolute values of the dephasing during two-year inspiral 
between the PN and the numerical waveforms for the dominant $\ell=m=2$ mode 
as a function of time in month. 
The left (right) panel represents the early (late) inspiral phase 
in the LISA band. 
The left panel shows the dephase for System-I of masses 
$(M,\m)=(10^5,10)M_{\odot}$, which evolves from 
$r_0\simeq 29M$ to $r_0\simeq 16M$ with associated frequencies 
$f_{\rm GW}\in [4\times 10^{-3},10^{-2}]$Hz. The right panel shows the dephase 
for System-II of masses $(M,\m)=(10^6,10)M_{\odot}$, 
which explores orbital radius 
in a region $r_0/M\in [6.0,11]$ and frequencies in a range 
$f_{\rm GW}\in [1.8\times 10^{-3},4.4\times 10^{-3}]$Hz. 
Note that the dephase between the 18PN (22PN) waveforms and numerical waveforms 
for System-I due to the two-year inspiral is about $8\times 10^{-9}$ 
($10^{-9}$) rads, which is below the lowest value of the dephase 
in the left panel. 
The dephase between the 22PN waveforms and numerical waveforms 
for System-II after the two-year inspiral is about $10^{-2}$ rads, which 
may suggest that the 22PN waveforms will provide the data analysis accuracy  
comparable to the one using numerical waveforms. 
}\label{fig:dephase22}
\end{center}
\end{figure}

Fig.~\ref{fig:dephase22} shows the absolute values of the phase difference 
of the dominant mode $h_{2,2}$ between the PN and the numerical calculation 
during the two-year inspiral. 
For System-I (II), the absolute values of the dephasing 
between the PN waveforms and the numerical waveforms 
after the two-year inspiral are about $4\times 10^{1}$ ($3\times 10^{3}$), 
$9\times 10^{-3}$ ($5\times 10^{1}$), 
$10^{-5}$ ($1$), $8\times 10^{-9}$ ($2\times 10^{-2}$) and 
$10^{-9}$ ($10^{-2}$) rads 
for 5.5PN, 10PN, 14PN, 18PN and 22PN respectively. 
Using the factorized resummation waveforms, the absolute values of 
the phase difference for System-I (II) can be reduced to 
$3\times 10^{-8}$ ($9\times 10^{-4}$) rads for 14PN waveforms~\cite{14PN} and 
$10^{-9}$ ($10^{-4}$) rads for our 22PN waveforms. 

Since System-II represents the inspiral in the most strong-field 
of a Schwarzschild black hole, 
the dephase between the 22PN waveforms and the numerical waveforms 
after two-year inspirals is expected to be smaller than $10^{-2}$ rads 
for extreme mass ratio binaries in the frequency band of LISA. 
This may indicate that 
the 22PN waveforms will lead to the accuracy of the data analysis of EMRIs 
comparable to the one resulting from numerical waveforms. 

%%%%%%%%%%%%%%%%%%%
%%%%%%%%%%%%%%%%%%%%%%%%%%%%%%%%%%%%%%%%%%%%%%%%%%%%%%%%%%%%%%%%
\subsection{Hybrid formula of the energy flux to infinity for a Kerr black hole}
\label{sec:flux_kerr}
\begin{figure}[htbp]
\begin{center}
\includegraphics[width=69mm]{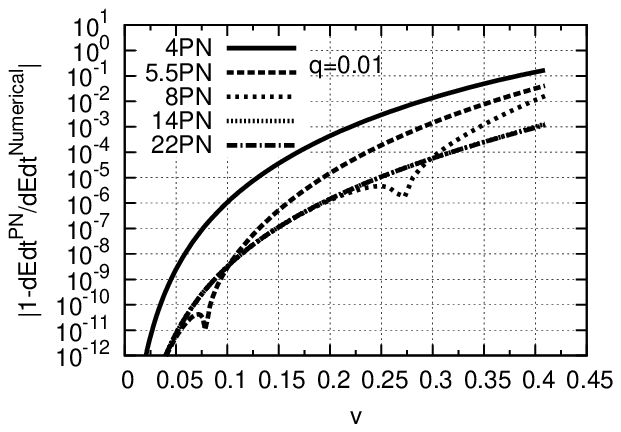}%
\includegraphics[width=69mm]{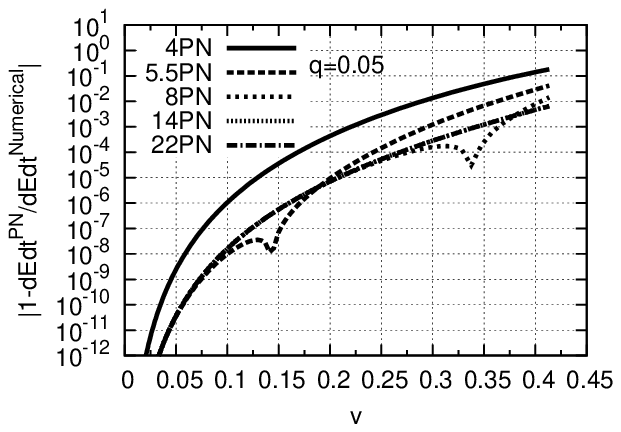}\\
\includegraphics[width=69mm]{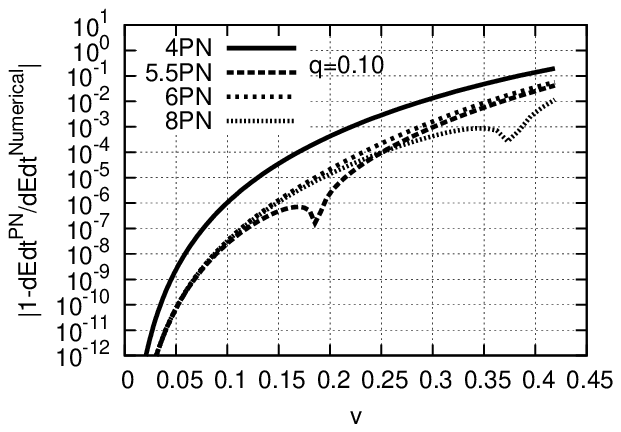}%
\includegraphics[width=69mm]{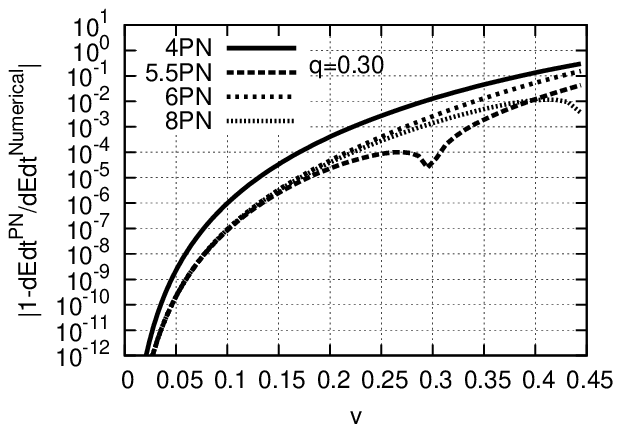}\\
\includegraphics[width=69mm]{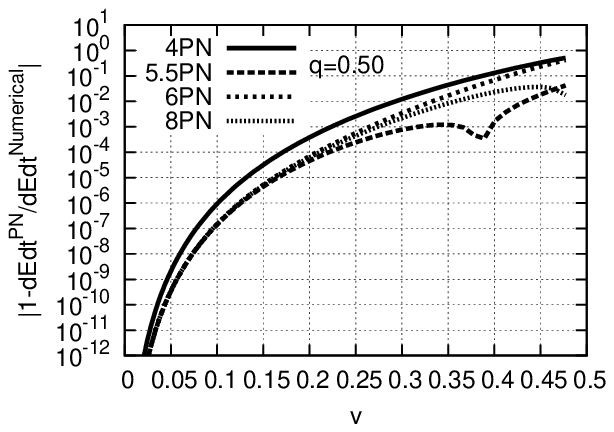}%
\includegraphics[width=69mm]{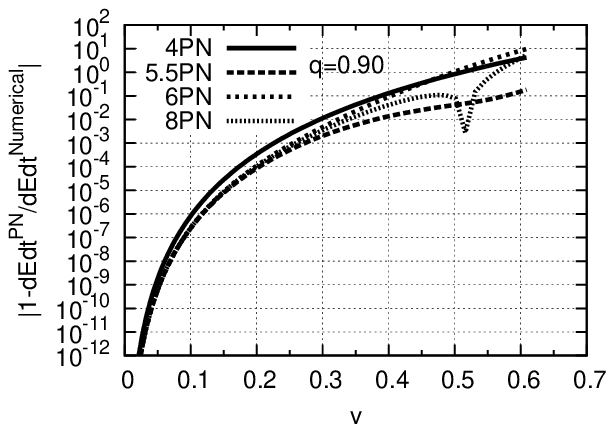}
\end{center}
\caption{
Absolute values of the difference of energy flux 
to infinity between numerical results and PN approximation 
as a function of orbital velocity, $v=(M/r_0)^{1/2}[1+q\,(M/r_0)^{3/2}]^{-1/3}$, 
up to ISCO for 
$q=0.01,\,0.05,\,0.1,\,0.3,\,0.5$ and $0.9$. 
The energy flux at $n$-PN combines 
$n$-PN energy flux for a Schwarzschild black hole and spin dependent terms 
of the 4PN energy flux for a Kerr black hole~\cite{ref:TSTS}. 
For $q<0.1$, higher PN order terms for a non-spinning black hole 
improve the relative accuracy of the energy flux. 
For $q\ge 0.1$, however, higher PN order terms for a non-spinning 
black hole do not always improve the accuracy although one can find 
the accuracy of some PN order is an order of magnitude better 
than that of the 4PN energy flux.
}\label{fig:flux_kerr}
\end{figure}

\begin{figure}[htbp]
\begin{center}
\includegraphics[width=69mm]{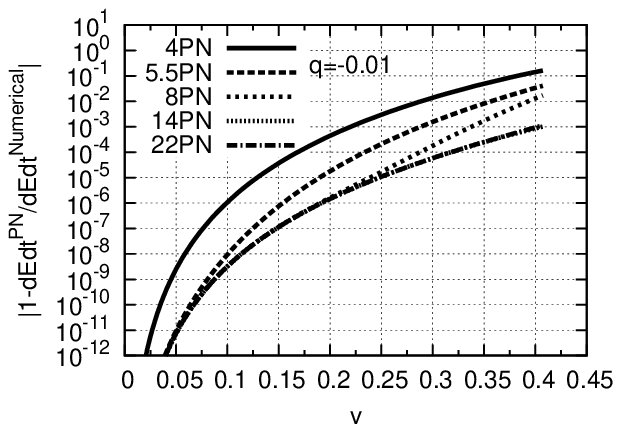}%
\includegraphics[width=69mm]{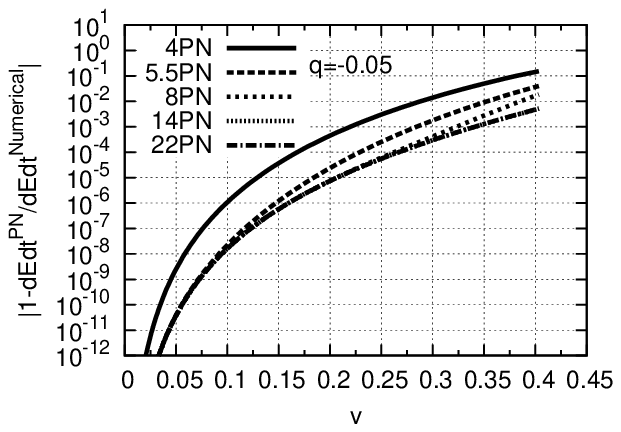}\\
\includegraphics[width=69mm]{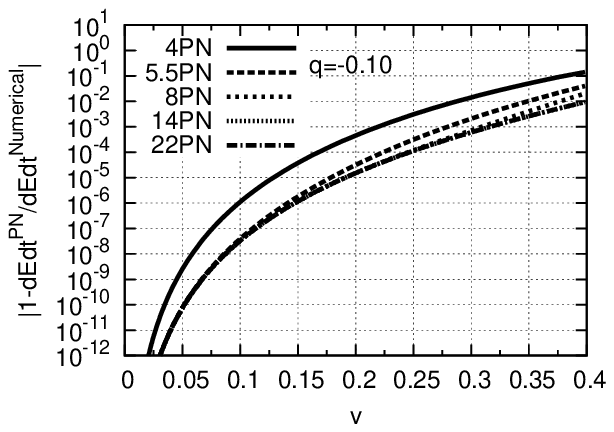}%
\includegraphics[width=69mm]{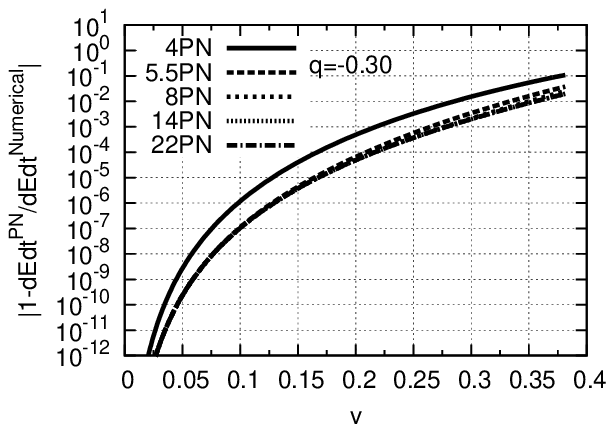}\\
\includegraphics[width=69mm]{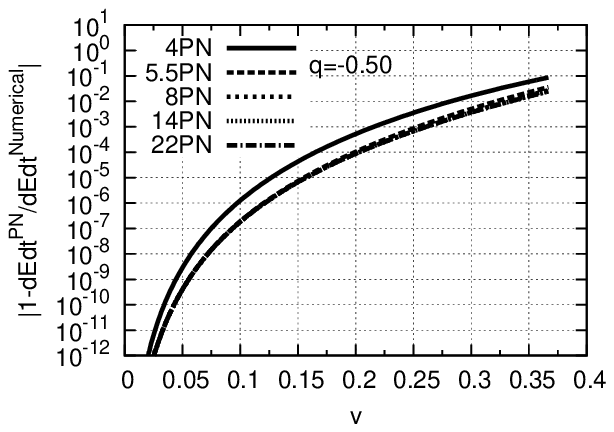}%
\includegraphics[width=69mm]{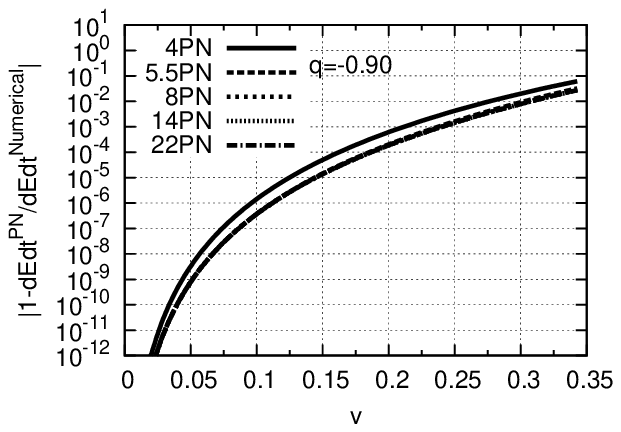}
\end{center}
\caption{
Same as Fig.~\ref{fig:flux_kerr} but for 
$q=-0.01,\,-0.05,\,-0.1,\,-0.3,\,-0.5$ and $q=-0.9$. 
For $|q|<0.1$, adding higher post-Newtonian order terms 
for a non-spinning black hole achieves about an order of magnitude 
better accuracy than using the 4PN energy flux for a spinning 
black hole. For $q<-0.1$ the improvement becomes small when $|q|$ large. 
}\label{fig:flux_kerr2}
\end{figure}

%%%%%%%%%%%%%%%%%%%%%%%%%%%%%%%%%%%%%%%%%%%%%%%%%%%%%%%%%%%%%%%%
To investigate how higher post-Newtonian order terms 
for a non-spinning black hole improve the accuracy of the 4PN formula 
for a spinning black hole~\cite{ref:TSTS}, 
we construct a hybrid formula for the energy flux 
in the post-Newtonian approximation by supplementing 
the 4PN energy flux for a Kerr black hole with 
our 22PN energy flux for a Schwarzschild black hole. 
In Figs.~\ref{fig:flux_kerr} and \ref{fig:flux_kerr2}, 
$n$-PN energy flux includes $n$-PN energy flux to infinity 
for a Schwarzschild black hole and spin dependent terms of the 4PN expression 
for the energy flux to infinity for a Kerr black hole. 
The numerical flux for a Kerr black hole is computed using 
the same technique~\cite{FT1,FT2} used in Sec.~\ref{sec:flux}. 
For the numerical calculation, we set $\ell=30$ which 
gives the relative error better than $10^{-5}$ up to ISCO for the 
parameters $q$ used in Figs.~\ref{fig:flux_kerr} and \ref{fig:flux_kerr2}. 

In Fig.~\ref{fig:flux_kerr} (Fig.~\ref{fig:flux_kerr2}), 
the comparisons for $q>0$ ($q<0$) are shown. From these figures, 
one will find that for $|q|<0.1$ adding higher post-Newtonian order terms 
for a non-spinning black hole achieves the accuracy that is about 
an order of magnitude better than the one using 4PN expression of 
the energy flux for a spinning black hole. 
For $|q|\ge 0.1$, however, the improvement saturates 
when the PN order is higher than 8PN. 
For $q<-0.1$ the improvement becomes small when $|q|$ large. 
For $q=-0.9$ one finds at most a factor of two improvement by 
adding the higher order terms for a non-spinning black hole. 
For $q=0.05$, the relative errors of the energy flux around ISCO are 
$2\times 10^{-1}$, $4\times 10^{-2}$ and $6\times 10^{-3}$ for 4PN, 5.5PN and 
8PN respectively. For $q=0.05$, adding 8.5PN or higher order terms does not 
achieve better accuracy than adding 8PN terms, but achieves better accuracy 
than adding 5.5PN terms. 
For $q\geq 0.1$, adding higher post-Newtonian order terms for a non-spinning 
black hole does not always improve the accuracy. 
For $q=0.1$, the relative errors of the 6PN energy flux around ISCO 
becomes larger than that of the 5.5PN energy flux around ISCO, but smaller 
than that of the 4PN energy flux around ISCO. 
For $q=0.5$, the relative error of the 6PN energy flux around ISCO is 
almost same as that of the 4PN energy flux around ISCO. 
For $q=0.9$, the relative error of the 6PN (5.5PN) energy flux around ISCO is 
two times larger (an order of magnitude smaller) than that of the 4PN energy 
flux. 

%%%%%%%%%%%%%%%%%%%
%%%%%%%%%%%%%%%%%%%

%%%%%%%%%%%%%%%%%%%
%%%%%%%%%%%%%%%%%%%%%%%%%%%%%%%%%%%%%%%%%%%%%%%%%%%%%%%%%%%%%%%%
\section{Summary}
\label{sec:summary}
%%%%%%%%%%%%%%%%%%%%%%%%%%%%%%%%%%%%%%%%%%%%%%%%%%%%%%%%%%%%%%%%
Using a systematic method to compute homogeneous solutions of the 
Teukolsky equation~\cite{ref:MST,ref:MSTR}, we have derived 
the 22PN expressions of gravitational waves 
for a test particle in circular orbits around a Schwarzschild black hole. 
The comparison of the energy flux between the PN expansion and high 
precision numerical results shows that the relative error of 
the 22PN energy flux around ISCO is about $10^{-5}$. 
We also find that the dephase between the 22PN waveforms and the numerical 
waveforms after two-year inspiral will be smaller than $10^{-2}$ rads 
for the most of the parameter space of EMRIs. 
This may imply that the 22PN waveforms will provide 
the accuracy in LISA-type data analysis comparable to the one using 
high precision numerical waveforms. (See Ref.~\citen{14PN} for 
the same discussion using factorized resummed waveforms, 
developed in Ref.~\citen{DIN}, at 14PN.) 

The next application will be the case in which 
a particle moves in circular orbits around a Kerr black hole. 
The current available result for the case is 4PN\cite{ref:TSTS}, which gives 
the relative errors of $10^{-1}$ ($O(1)$) for $q=0.1$ ($q=0.9$) 
around ISCO (See Figs.~\ref{fig:flux_kerr} and \ref{fig:flux_kerr2}). 
To investigate whether our 22PN formula of the energy flux improves 
the accuracy for the case of a Kerr black hole, 
we construct a hybrid formula of the energy flux by 
supplementing the 4PN energy flux for a Kerr black hole with 
the 22PN energy flux for a Schwarzschild black hole. 
We found that for $q<0.1$ non-spinning terms of higher post-Newtonian order 
expressions improve the accuracy around ISCO. 
However, since for $q\geq 0.1$ non-spinning terms of higher 
post-Newtonian order expressions 
of the total energy flux do not always improve the accuracy for a large 
value of the spin, 
it is necessary to obtain high PN order expressions 
for the case of a Kerr black hole. 
For $q=0.9$, the numerical calculation shows that one needs to include 
$\ell=30$ modes to achieve the relative error of the total energy flux 
as $10^{-5}$ around ISCO (See Sec.~\ref{sec:flux_kerr}). 
This may indicate that we need to compute at least up to 28PN to 
obtain the relative error of $10^{-5}$ around ISCO for $q=0.9$. 
It may be difficult to perform 
such a high post-Newtonian order calculation using our current code 
since the number of terms necessary to derive the PN expressions 
grows exponentially when the PN order becomes higher 
(See Sec.~\ref{sec:intro} and Ref.~\citen{14PN}). 
If one can not obtain sufficiently high post-Newtonian order expressions, 
another approach may be the effective-one-body formalism which 
can include unknown coefficients in the post-Newtonian expressions 
by calibrating them with numerical results~\cite{EOB_EMRI,YBHPBMT}. 

%%%%%%%%%%%%%%%%%%%%%%%%%%%%%%%%%%%%%%%%%%%%%%%%%%%%%%%%%%%%%%%%
\section*{Acknowledgements}

It is a pleasure to thank Bala R. Iyer for continuous encouragement 
to look into this problem and useful comments on the manuscript. 
The author is grateful for the support of the European Union FEDER funds, 
the Spanish Ministry of Economy and Competitiveness (projects 
FPA2010-16495 and CSD2007-00042) and the Conselleria 
d'Economia Hisenda i Innovacio of the Govern de les Illes Balears. 
%%%%%%%%%%%%%%%%%%%%%%%%%%%%%%%%%%%%%%%%%%%%%%%%%%%%%%%%%%%%%%%%
%   Appendix
%%%%%%%%%%%%%%%%%%%%%%%%%%%%%%%%%%%%%%%%%%%%%%%%%%%%%%%%%%%%%%%%
\appendix
%%%%%%%%%%%%%%%%%%%%%%%%%%%%%%%%%%%%%%%%%%%%%%%%%%%%%%%%%%%%%%%%
\section{7PN formula for the energy flux to infinity} 
\label{sec:7pn_formula}

The total energy flux to infinity at 7PN becomes as
\begin{eqnarray*}
{dE\over dt}=&&\left({dE\over dt}\right)_N \Biggl[
1
-{\displaystyle \frac {1247}{336}} v^2
+4\,\pi v^3
-{\displaystyle \frac {44711}{9072}} v^4
-{\displaystyle \frac {8191}{672}} \,\pi v^5 
\cr&&\cr&&\vspace{6pt}
+\left\{
{\displaystyle \frac {6643739519}{69854400}}  - 
{\displaystyle \frac {1712}{105}} \,\gamma  - {\displaystyle 
\frac {3424}{105}} \,\ln(2) + {\displaystyle \frac {16}{3
}} \,\pi ^{2} - {\displaystyle \frac {1712}{105}} \,\ln(v
)
\right\} v^6
- {\displaystyle \frac {16285}{504}} \,\pi v^7 
\cr&&\cr&&\vspace{6pt}
+\left\{
 - {\displaystyle \frac {323105549467}{3178375200}}  + 
{\displaystyle \frac {232597}{4410}} \,\gamma  + {\displaystyle 
\frac {39931}{294}} \,\ln(2) - {\displaystyle \frac {1369
}{126}} \,\pi ^{2} - {\displaystyle \frac {47385}{1568}} \,
\ln(3) \right.\cr&&\cr&&\left.\vspace{6pt}\hspace{0.6cm}
 + {\displaystyle \frac {232597}{4410}} \,\ln(v)
\right\} v^8 \cr&&\cr&&\vspace{6pt}
+\left\{
{\displaystyle \frac {265978667519}{745113600}} \,\pi  - 
{\displaystyle \frac {13696}{105}} \,\ln(2)\,\pi  - 
{\displaystyle \frac {6848}{105}} \,\pi \,\gamma  - 
{\displaystyle \frac {6848}{105}} \,\pi \,\ln(v)
\right\} v^9 \cr&&\cr&&\vspace{6pt}
+\left\{
 - {\displaystyle \frac {2500861660823683}{2831932303200
}}  + {\displaystyle \frac {916628467}{7858620}} \,\gamma  - 
{\displaystyle \frac {83217611}{1122660}} \,\ln(2) - 
{\displaystyle \frac {424223}{6804}} \,\pi ^{2} 
\right.\cr&&\cr&&\left.\vspace{6pt}\hspace{0.6cm}
 + {\displaystyle \frac {47385}{196}} \,\ln(3) + 
{\displaystyle \frac {916628467}{7858620}} \,\ln(v)
\right\} v^{10}\cr&&\cr&&\vspace{6pt}
+\left\{
{\displaystyle \frac {177293}{1176}} \,\pi \,\gamma  + 
{\displaystyle \frac {8521283}{17640}} \,\ln(2)\,\pi  + 
{\displaystyle \frac {8399309750401}{101708006400}} \,\pi  - 
{\displaystyle \frac {142155}{784}} \,\pi \,\ln(3) 
 \right.\cr&&\cr&&\left.\vspace{6pt}\hspace{0.6cm}
 + {\displaystyle \frac {177293}{1176}} \,\pi \,{\rm ln}(v)
\right\} v^{11}
\cr&&\cr&&\vspace{6pt}
+\left\{
 - {\displaystyle \frac {256}{45}} \,\pi ^{4} - {\displaystyle \frac {37744140625}{260941824}} \,\ln(5) + {\displaystyle \frac {2067586193789233570693}{602387400044430000}}  
\right.\cr&&\cr&&\vspace{6pt}\hspace{0.6cm}
 - {\displaystyle \frac {246137536815857}{157329572400}} \,\gamma  - {\displaystyle \frac {27392}{105}} \,\zeta (3) - {\displaystyle \frac {437114506833}{789268480}} \,\ln(3)
 \cr&&\cr&&\vspace{6pt}\hspace{0.6cm}
 - {\displaystyle \frac {271272899815409}{157329572400}} \,\ln(2) + {\displaystyle \frac {5861888}{11025}} \,\ln(2)\,\gamma  - {\displaystyle \frac {54784}{315}} \,\ln(2)\,\pi ^{2} 
\cr&&\cr&&\vspace{6pt}\hspace{0.6cm}
 + {\displaystyle \frac {3803225263}{10478160}} \,\pi ^{2} - {\displaystyle \frac {27392}{315}} \,\pi ^{2}\,\gamma  + {\displaystyle \frac {5861888}{11025}} \,\ln(2)^{2} + {\displaystyle \frac {1465472}{11025}} \,\gamma ^{2} 
\cr&&\cr&&\vspace{6pt}\hspace{0.6cm}
 + \left({\displaystyle \frac {2930944}{11025}} \,\gamma  - {\displaystyle \frac {27392}{315}} \,\pi ^{2} - {\displaystyle \frac {246137536815857}{157329572400}}  + {\displaystyle \frac {5861888}{11025}} \,\ln(2)\right)\,\ln(v) 
\cr&&\cr&&\left.\vspace{6pt}\hspace{0.6cm}
 + {\displaystyle \frac {1465472}{11025}} \,\ln(v)^{2} 
\right\} v^{12} \cr&&\cr&&\vspace{6pt}
+\left\{
{\displaystyle \frac {300277177}{436590}} \,\pi \,\gamma  - {\displaystyle \frac {81605095538444363}{20138185267200}} \,\pi  - {\displaystyle \frac {42817273}{71442}} \,\ln(2)\,\pi  
\right.\cr&&\cr&&\left.\vspace{6pt}\hspace{0.6cm}
 + {\displaystyle \frac {142155}{98}} \,\pi \,\ln(3) + {\displaystyle \frac {300277177}{436590}} \,\pi \,{\rm ln}(v)
\right\} v^{13} \cr&&\cr&&\vspace{6pt}
+\left\{
{\displaystyle \frac {531077}{2205}} \,\zeta (3) + {\displaystyle \frac {19402232550751339}{17896238860500}} \,\ln(2) 
\right.\cr&&\cr&&\vspace{6pt}\hspace{0.6cm}
 + {\displaystyle \frac {58327313257446476199371189}{8332222517414555760000}}  + {\displaystyle \frac {128223}{245}} \,\ln(2)\,\pi ^{2} - {\displaystyle \frac {9523}{945}} \,\pi ^{4} 
\cr&&\cr&&\vspace{6pt}\hspace{0.6cm}
 - {\displaystyle \frac {5811697}{2450}} \,\ln(2)^{2} + {\displaystyle \frac {1848015}{2744}} \,\gamma \,{\rm ln}(3) - {\displaystyle \frac {471188717}{231525}} \,\ln(2)\,\gamma  
\cr&&\cr&&\vspace{6pt}\hspace{0.6cm}
 - {\displaystyle \frac {6136997968378863}{1256910054400}} \,\ln(3) + {\displaystyle \frac {9926708984375}{5088365568}} \,\ln(5) 
\cr&&\cr&&\vspace{6pt}\hspace{0.6cm} 
+ {\displaystyle \frac {1848015}{2744}} \,\ln(2)\,\ln(3) - {\displaystyle \frac {142155}{392}} \,{\rm ln}(3)\,\pi ^{2} + {\displaystyle \frac {9640384387033067}{17896238860500}} \,\gamma  
\cr&&\cr&&\vspace{6pt}\hspace{0.6cm}
 - {\displaystyle \frac {52525903}{154350}} \,\gamma ^{2} + {\displaystyle \frac {531077}{6615}} \,\pi ^{2}\,\gamma  + {\displaystyle \frac {2621359845833}{2383781400}} \,\pi ^{2} + {\displaystyle \frac {1848015}{5488}} \,\ln(3)^{2} 
 \cr&&\cr&&\vspace{6pt}\hspace{0.6cm}
+ \left({\displaystyle \frac {9640384387033067}{17896238860500}}  - {\displaystyle \frac {471188717}{231525}} \,\ln(2) + {\displaystyle \frac {531077}{6615}} \,\pi ^{2} - {\displaystyle \frac {52525903}{77175}} \,\gamma  \right.
\cr&&\cr&&\left.\vspace{6pt}\hspace{0.6cm}
\left. + {\displaystyle \frac {1848015}{2744}} \,\ln(3)\right)\ln(v) 
- {\displaystyle \frac {52525903}{154350}} \,\ln(v)^{2} 
\right\} v^{14}\Biggr],
\end{eqnarray*}
where $\gamma$ is the Euler constant and $\zeta(n)$ is the Zeta function. 

We note that coefficients in the post-Newtonian expansion for the total 
energy flux are polynomial functions of $\pi$, $\gamma$, $\ln(n)$, $\ln(v)$ 
and $\zeta(n)$. One will find that the coefficient at 1.5PN, $4\pi$, 
is derived from the PN expansion of $e^{-\pi\epsilon/2}$ in Eq.~(\ref{eq:Ap}) 
for $\ell=m=2$ mode. One will also find that 
$\gamma$, $\ln(2)$ and $\ln(v)$ terms at 3PN and $\zeta(3)$ term at 6PN 
are derived by the PN expansion of the homogeneous Teukolsky solution 
in a series of Coulomb wave functions Eq.~(\ref{eq:series of Rc})
for $\ell=m=2$ mode. Similarly, one will find that 
$\ln(n)$ and $\zeta(2n-1)$ terms appear from $(n+1)$-PN and 
$3n$-PN respectively (See Ref.~\citen{BHPC}). 
We also note that $(\ln (v))^n$ terms appear from $3n$-PN 
(See Appendix~\ref{sec:22pn_formula} and Ref.~\citen{BHPC}).
One of the reasons is that $(\ln (v))^n$ terms at 3$n$-PN are produced by 
the PN expansion of $z^{\nu}$ in the homogeneous Teukolsky solution 
in a series of Coulomb wave functions Eq.~(\ref{eq:series of Rc}), 
where $z=\omega r=O(v)$ and 
$\nu=\ell+\nu^{(2)}\,v^6+O(v^9)$. Noting that the energy flux is computed 
by the square of the homogeneous Teukolsky solution and $\nu^{(2)}=-856/105$ for 
the dominant mode $\ell=m=2$~\cite{ref:MST,ref:MSTR,ST}, one can estimate 
leading $(\ln (v))^n$ terms in the energy flux using series expansion of 
$v^{-1712/105\,v^6}$ and find that these $(\ln (v))^n$ terms at 3$n$-PN 
agree with the ones in our 22PN energy flux. 
The explicit expressions for these leading $(\ln (v))^n$ terms 
at 3$n$-PN are already derived in Eq.~(44) of Ref.~\citen{ref:GR2010} 
using the renormalization group equations and agree with the ones 
in our 22PN energy flux. 
%%%%%%%%%%%%%%%%%%%%%%%%%%%%%%%%%%%%%%%%%%%%%%%%%%%%%%%%%%%%%%%%
\section{22PN energy flux to infinity for numerical calculation} 
\label{sec:22pn_formula}

The total energy flux to infinity at 22PN, which can be used for 
numerical calculation for double precision calculation, becomes as
\begin{eqnarray*}
{dE\over dt}=&&\left({dE\over dt}\right)_N \Biggl[
1 - 3.7113095238095238\,{v}^{2} +  12.566370614359173\,{v}^{3}
\cr&&\hspace{1.5cm}
- 4.9284611992945326\,{v}^{4}
- 38.292835454693446\,{v}^{5}
\cr &&\cr &&\hspace{1.5cm}\vspace{2pt}
+\left\{ 
115.73171667561132- 16.304761904761905\,\ln (v) 
\right\}{v}^{6}
\cr &&\cr &&\hspace{1.5cm}\vspace{2pt}
- 101.50959595974163\,{v}^{7}
\cr &&\cr &&\hspace{1.5cm}\vspace{2pt}
+\left\{ 
- 117.50439072267733+ 52.743083900226757\,\ln (v) 
\right\}{v}^{8}
\cr &&\cr &&\hspace{1.5cm}\vspace{2pt}
+\left\{ 
 719.12834223342969- 204.89168087412289\,\ln  ( v ) 
\right\}{v}^{9}
\cr &&\cr &&\hspace{1.5cm}\vspace{2pt}
+\left\{ 
- 1216.9069913170420+ 116.63987659410940\,\ln  ( v ) 
\right\}{v}^{10}
\cr &&\cr &&\hspace{1.5cm}\vspace{2pt}
+\left\{ 
 958.93497011956684+ 473.62447817423062\,\ln  ( v ) 
\right\}{v}^{11}
\cr &&\cr &&\hspace{1.5cm}\vspace{2pt}
+\left\{ 
 2034.7899821303756- 1900.7293251880964\,\ln  ( v ) 
\right.\cr &&\cr &&\left.\hspace{1.5cm}\vspace{2pt}
 + 132.92263038548753\, \ln  (v) ^{2}
\right\}{v}^{12}
\cr &&\cr &&\hspace{1.5cm}\vspace{2pt}
+\left\{ 
- 7781.9735576745461+ 2160.7196071918322\,\ln  ( v ) 
\right\}{v}^{13}
\cr &&\cr &&\hspace{1.5cm}\vspace{2pt}
+\left\{ 
 15384.154335011960+ 267.42096713731240\,\ln  ( v ) 
\right.\cr &&\cr &&\left.\hspace{1.5cm}\vspace{2pt}
 - 340.30387431162941\, \ln  ( v ) ^{2}
\right\}{v}^{14}
\cr &&\cr &&\hspace{1.5cm}\vspace{2pt}
+\left\{ 
- 12154.148413021627- 10431.406072327709\,\ln  ( v ) 
\right.\cr &&\cr &&\left.\hspace{1.5cm}\vspace{2pt}
+ 1670.3550364595162\, \ln  ( v ) ^{2}
\right\}{v}^{15}
\cr &&\cr &&\hspace{1.5cm}\vspace{2pt}
+\left\{ 
- 14423.548835921780+ 21798.970641540720\,\ln  ( v ) 
\right.\cr &&\cr &&\left.\hspace{1.5cm}\vspace{2pt}
- 1259.2120782299356\, \ln  ( v )  ^{2}
\right\}{v}^{16}
\cr &&\cr &&\hspace{1.5cm}\vspace{2pt}
+\left\{ 
 89501.573309687739- 29246.578925458487\,\ln  ( v ) 
\right.\cr &&\cr &&\left.\hspace{1.5cm}\vspace{2pt}
- 2162.0533013852443\, \ln  ( v ) ^{2}
\right\}{v}^{17}
\cr &&\cr &&\hspace{1.5cm}\vspace{2pt}
+\left\{ 
- 183583.88710261773- 6836.8321880197775\,\ln  ( v ) 
\right.\cr &&\cr &&\left.\hspace{1.5cm}\vspace{2pt}
+ 14856.770074354592\, \ln  ( v ) ^{2}- 722.42394673001476\, \ln  ( v ) ^{3}
\right\}{v}^{18}
\cr &&\cr &&\hspace{1.5cm}\vspace{2pt}
+\left\{ 
 232409.13476410614+ 94495.266713306094\,\ln  ( v ) 
\right.\cr &&\cr &&\left.\hspace{1.5cm}\vspace{2pt}
- 20805.243844022191\, \ln  ( v ) ^{2}
\right\}{v}^{19}
\cr &&\cr &&\hspace{1.5cm}\vspace{2pt}
+\left\{ 
 29936.08409119159- 292760.39953957798\,\ln  ( v ) 
\right.\cr &&\cr &&\left.\hspace{1.5cm}\vspace{2pt}
+ 16701.438906657108\, \ln  ( v ) ^{2}+ 1179.1301184451609\, \ln  ( v ) ^{3}
\right\}{v}^{20}
\cr &&\cr &&\hspace{1.5cm}\vspace{2pt}
+\left\{ 
- 825753.06115027078+ 441939.66010505905\,\ln  ( v ) 
\right.\cr &&\cr &&\left.\hspace{1.5cm}\vspace{2pt}
+ 57856.092251923344\, \ln  ( v ) ^{2}- 9078.2470552974338\, \ln  ( v ) ^{3}
\right\}{v}^{21}
\cr &&\cr &&\hspace{1.5cm}\vspace{2pt}
+\left\{ 
 2458416.5416899955- 398672.73444594275\,\ln  ( v ) 
\right.\cr &&\cr &&\left.\hspace{1.5cm}\vspace{2pt}
- 158992.86702314628\, \ln  ( v ) ^{2}+ 8194.4543122810763\, \ln  ( v )^{3}
\right\}{v}^{22}
\cr &&\cr &&\hspace{1.5cm}\vspace{2pt}
+\left\{ 
- 3952774.9138307063- 759261.92414463453\,\ln  ( v )
\right.\cr &&\cr &&\left.\hspace{1.5cm}\vspace{2pt}
 + 360084.58553892898\, \ln  ( v ) ^{2}- 897.89185596522682\, \ln  ( v ) ^{3}
\right\}{v}^{23}
\cr &&\cr &&\hspace{1.5cm}\vspace{2pt}
+\left\{ 
 2976918.2174426533+ 3500957.1806440358\,\ln  ( v ) 
\right.\cr &&\cr &&\left.\hspace{1.5cm}\vspace{2pt}
- 318467.34660011464\, \ln  ( v ) ^{2}- 70899.505625286470\, \ln  ( v ) ^{3}
\right.\cr &&\cr &&\left.\hspace{1.5cm}\vspace{2pt}
+ 2944.7376114328221\, \ln  ( v ) ^{4}
\right\}{v}^{24}
\cr &&\cr &&\hspace{1.5cm}\vspace{2pt}
+\left\{ 
 7019236.3677905130- 8886564.3632113352\,\ln  ( v ) 
\right.\cr &&\cr &&\left.\hspace{1.5cm}\vspace{2pt}
+ 52809.58085577387\, \ln  ( v ) ^{2}+ 113193.57722837660\, \ln  ( v ) ^{3}
\right\}{v}^{25}
\cr &&\cr &&\hspace{1.5cm}\vspace{2pt}
+\left\{ 
- 31202976.107218423+ 11762892.421868797\,\ln  ( v ) 
\right.\cr &&\cr &&\left.\hspace{1.5cm}\vspace{2pt}
+ 1982484.3746860539\, \ln  ( v ) ^{2}- 220750.96927550645\, \ln  ( v ) ^{3}
\right.\cr &&\cr &&\left.\hspace{1.5cm}\vspace{2pt}
- 1067.0133724433031\, \ln  ( v ) ^{4}
\right\}{v}^{26}
\cr &&\cr &&\hspace{1.5cm}\vspace{2pt}
+\left\{ 
 68807279.901880549- 2488093.498633921\,\ln  ( v ) 
\right.\cr &&\cr &&\left.\hspace{1.5cm}\vspace{2pt}
- 5652341.0354988102\, \ln  ( v ) ^{2}- 45136.07364771081\, \ln  ( v )^{3}
\right.\cr &&\cr &&\left.\hspace{1.5cm}\vspace{2pt}
+ 37004.664187307634\, \ln  ( v ) ^{4}
\right\}{v}^{27}
\cr &&\cr &&\hspace{1.5cm}\vspace{2pt}
+\left\{ 
- 79540420.970120669- 47212469.571473335\,\ln  ( v ) 
\right.\cr &&\cr &&\left.\hspace{1.5cm}\vspace{2pt}
+ 13096024.087086880\, \ln  ( v ) ^{2}+ 336827.97427818428\, \ln  ( v ) ^{3}
\right.\cr &&\cr &&\left.\hspace{1.5cm}\vspace{2pt}
- 34129.747458105155\, \ln  ( v ) ^{4}
\right\}{v}^{28}
\cr &&\cr &&\hspace{1.5cm}\vspace{2pt}
+\left\{ 
- 25925436.707911899+ 148981108.02576149\,\ln  ( v ) 
\right.\cr &&\cr &&\left.\hspace{1.5cm}\vspace{2pt}
- 12498185.772224907\, \ln  ( v ) ^{2}- 2301205.5277433290\, \ln  ( v ) ^{3}
\right.\cr &&\cr &&\left.\hspace{1.5cm}\vspace{2pt}
+ 74157.041397297642\, \ln  ( v ) ^{4}
\right\}{v}^{29}
\cr &&\cr &&\hspace{1.5cm}\vspace{2pt}
+\left\{ 
 402444362.19171711- 278017845.54926934\,\ln  ( v ) 
\right.\cr &&\cr &&\left.\hspace{1.5cm}\vspace{2pt}
- 11687734.115250883\, \ln  ( v ) ^{2}+ 4307818.2864565577\, \ln  ( v ) ^{3}
\right.\cr &&\cr &&\left.\hspace{1.5cm}\vspace{2pt}
+ 213688.37579621078\, \ln  ( v )  ^{4}- 9602.6491252818883\, \ln  ( v ) ^{5}
\right\}{v}^{30}
\cr &&\cr &&\hspace{1.5cm}\vspace{2pt}
+\left\{ 
- 1071349694.3775396+ 204222856.19303166\,\ln  ( v ) 
\right.\cr &&\cr &&\left.\hspace{1.5cm}\vspace{2pt}
+ 116030090.57483748\, \ln  ( v ) ^{2}- 10073180.152647150\, \ln  ( v ) ^{3}
\right.\cr &&\cr &&\left.\hspace{1.5cm}\vspace{2pt}
- 291438.02818774543\, \ln  ( v ) ^{4}
\right\}{v}^{31}
\cr &&\cr &&\hspace{1.5cm}\vspace{2pt}
+\left\{ 
 1626083110.9818554+ 488335019.02563548\,\ln  ( v ) 
\right.\cr &&\cr &&\left.\hspace{1.5cm}\vspace{2pt}
- 291331989.87961219\, \ln  ( v ) ^{2}+ 2942116.4939216327\, \ln  ( v ) ^{3}
\right.\cr &&\cr &&\left.\hspace{1.5cm}\vspace{2pt}
+ 1469109.8755678809\, \ln  ( v ) ^{4}- 13190.813005345736\, \ln  ( v ) ^{5}
\right\}{v}^{32}
\cr &&\cr &&\hspace{1.5cm}\vspace{2pt}
+\left\{ 
- 544918059.05664234- 2428299657.0865688\,\ln  ( v )
\right.\cr &&\cr &&\left.\hspace{1.5cm}\vspace{2pt}
 + 456879101.34472726\, \ln  ( v ) ^{2}+ 26594005.608730906\, \ln  ( v ) ^{3}
\right.\cr &&\cr &&\left.\hspace{1.5cm}\vspace{2pt}
- 1380784.1214801066\, \ln  ( v ) ^{4}- 120670.44778794414\, \ln  ( v ) ^{5}
\right\}{v}^{33}
\cr &&\cr &&\hspace{1.5cm}\vspace{2pt}
+\left\{ 
- 4616517093.2988715+ 5243279919.6114690\,\ln  ( v ) 
\right.\cr &&\cr &&\left.\hspace{1.5cm}\vspace{2pt}
- 38163582.107515614\, \ln  ( v ) ^{2}- 148107440.84908515\, \ln  ( v ) ^{3}
\right.\cr &&\cr &&\left.\hspace{1.5cm}\vspace{2pt}
+ 4207389.1083361737\, \ln  ( v ) ^{4}+ 75160.959023984080\, \ln  ( v ) ^{5}
\right\}{v}^{34}
\cr &&\cr &&\hspace{1.5cm}\vspace{2pt}
+\left\{ 
 15817926503.756157- 6113621574.3296087\,\ln  ( v ) 
\right.\cr &&\cr &&\left.\hspace{1.5cm}\vspace{2pt}
- 1759324394.0096811\, \ln  ( v ) ^{2}+ 301378311.00479927\, \ln  ( v ) ^{3}
\right.\cr &&\cr &&\left.\hspace{1.5cm}\vspace{2pt}
+ 5936380.6862923426\, \ln  ( v ) ^{4}- 556060.02405979572\, \ln  ( v ) ^{5}
\right\}{v}^{35}
\cr &&\cr &&\hspace{1.5cm}\vspace{2pt}
+\left\{ 
- 28150296022.364384- 3508984690.6951288\,\ln  ( v ) 
\right.\cr &&\cr &&\left.\hspace{1.5cm}\vspace{2pt}
+ 6000317955.561223\, \ln  ( v )  ^{2}- 384692002.27108097\, \ln  ( v ) ^{3}
\right.\cr &&\cr &&\left.\hspace{1.5cm}\vspace{2pt}
- 22068300.433686751\, \ln  ( v ) ^{4}- 341412.56352301983\, \ln  ( v )  ^{5}
\right.\cr &&\cr &&\left.\hspace{1.5cm}\vspace{2pt}
+ 26094.817940448560\, \ln  ( v ) ^{6}
\right\}{v}^{36}
\cr &&\cr &&\hspace{1.5cm}\vspace{2pt}
+\left\{ 
21923504375.507727+35127491396.323879\,\ln(v)
\right.\cr &&\cr &&\left.\hspace{1.5cm}\vspace{2pt}
-10561491802.698468\,\ln(v)^2-475718215.99686552\,\ln(v)^3
\right.\cr &&\cr &&\left.\hspace{1.5cm}\vspace{2pt}
+111153302.13498037\,\ln(v)^4-796270.98236252163\,\ln(v)^5
\right\}{v}^{37}
\cr &&\cr &&\hspace{1.5cm}\vspace{2pt}
+\left\{ 
45025396338.391247-92707192688.847145\,\ln(v)
\right.\cr &&\cr &&\left.\hspace{1.5cm}\vspace{2pt}
+7698249771.7266011\,\ln(v)^2+2977537585.1450740\,\ln(v)^3
\right.\cr &&\cr &&\left.\hspace{1.5cm}\vspace{2pt}
-169371453.19161573\,\ln(v)^4-6119765.2693920173\,\ln(v)^5
\right.\cr &&\cr &&\left.\hspace{1.5cm}\vspace{2pt}
+97766.114767012667\,\ln(v)^6
\right\}{v}^{38}
\cr &&\cr &&\hspace{1.5cm}\vspace{2pt}
+\left\{
-214410246975.35927+133503754160.33392\,\ln(v)
\right.\cr &&\cr &&\left.\hspace{1.5cm}\vspace{2pt}
+23775901926.064545\,\ln(v)^2-8155501690.3403790\,\ln(v)^3
\right.\cr &&\cr &&\left.\hspace{1.5cm}\vspace{2pt}
+154065244.97484914\,\ln(v)^4+9046986.8078241302\,\ln(v)^5
\right.\cr &&\cr &&\left.\hspace{1.5cm}\vspace{2pt}
+327917.15335390533\,\ln(v)^6
\right\}{v}^{39}
\cr &&\cr &&\hspace{1.5cm}\vspace{2pt}
+\left\{
457980639365.70258-27000183902.801376\,\ln(v)
\right.\cr &&\cr &&\left.\hspace{1.5cm}\vspace{2pt}
-105502717126.40338\,\ln(v)^2+10988802554.611972\,\ln(v)^3
\right.\cr &&\cr &&\left.\hspace{1.5cm}\vspace{2pt}
+812919382.11642567\,\ln(v)^4-51924388.655256156\,\ln(v)^5
\right.\cr &&\cr &&\left.\hspace{1.5cm}\vspace{2pt}
+120256.87755099946\,\ln(v)^6
\right\}{v}^{40}
\cr &&\cr &&\hspace{1.5cm}\vspace{2pt}
+\left\{
-514130514219.36581-465015514007.42288\,\ln(v)
\right.\cr &&\cr &&\left.\hspace{1.5cm}\vspace{2pt}
+229651541194.18049\,\ln(v)^2+80213128.550237347\,\ln(v)^3
\right.\cr &&\cr &&\left.\hspace{1.5cm}\vspace{2pt}
-2851170970.3905610\,\ln(v)^4+29259922.934207013\,\ln(v)^5
\right.\cr &&\cr &&\left.\hspace{1.5cm}\vspace{2pt}
+2678253.6091283655\,\ln(v)^6
\right\}{v}^{41}
\cr &&\cr &&\hspace{1.5cm}\vspace{2pt}
+\left\{
-246285643927.15478+1492672221227.6888\,\ln(v)
\right.\cr &&\cr &&\left.\hspace{1.5cm}\vspace{2pt}
-247860628603.14260\,\ln(v)^2-55696072359.371755\,\ln(v)^3
\right.\cr &&\cr &&\left.\hspace{1.5cm}\vspace{2pt}
+6850753270.3130100\,\ln(v)^4-23989989.755180755\,\ln(v)^5
\right.\cr &&\cr &&\left.\hspace{1.5cm}\vspace{2pt}
+52812.42772484907\,\ln(v)^6-60781.399066731884\,\ln(v)^7
\right\}{v}^{42}
\cr &&\cr &&\hspace{1.5cm}\vspace{2pt}
+\left\{
2724640759315.9973-2613409597570.6299\,\ln(v)
\right.\cr &&\cr &&\left.\hspace{1.5cm}\vspace{2pt}
-213879053529.96990\,\ln(v)^2+171699406240.25908\,\ln(v)^3
\right.\cr &&\cr &&\left.\hspace{1.5cm}\vspace{2pt}
-5303254029.5716242\,\ln(v)^4-666793776.26385442\,\ln(v)^5
\right.\cr &&\cr &&\left.\hspace{1.5cm}\vspace{2pt}
+13432837.504069723\,\ln(v)^6
\right\}{v}^{43}
\cr &&\cr &&\hspace{1.5cm}\vspace{2pt}
+\left\{
-7046976172838.4569+1802703723560.9900\,\ln(v)
\right.\cr &&\cr &&\left.\hspace{1.5cm}\vspace{2pt}
+1747504187818.0615\,\ln(v)^2-311569476082.28712\,\ln(v)^3
\right.\cr &&\cr &&\left.\hspace{1.5cm}\vspace{2pt}
-11182012302.436379\,\ln(v)^4+1624600040.7560485\,\ln(v)^5
\right.\cr &&\cr &&\left.\hspace{1.5cm}\vspace{2pt}
+15277409.847611050\,\ln(v)^6-424858.25453008154\,\ln(v)^7
\right\}{v}^{44}
\Biggr].
\end{eqnarray*}

%%%%%%%%%%%%%%%%%%%%%%%%%%%%%%%%%%%%%%%%%%%%%%%%%%%%%%%%%%%%%%%%%%%%%%%%%%%

%%%%%%%%%%%%%%%%%%%%%%%%%%%%%%%%%%%%%%%%%%%%%%%%%%%%%%%%%%%%%%%%
\end{document}